# All Recognition is Accomplished By Interacting Bottom-Up Sensory and Top-Down Context Biasing Occipital to Frontal Cortex Neural Networks

John S. Antrobus, Yusuke Shono, Wolfgang M. Pauli, and Bala Sundaram

## Abstract

Recognition of every word is accomplished by the close collaboration of bottom-up sub-word and word recognition neural networks with top-down neurocognitive word context expectations.  The utility of this context-appropriate collaboration is substantial savings in recognition time, accuracy and cortical neural processing resources.  Repetition priming, word recognition by prior reading of the same word, the simplest form of context facilitation, has been studied extensively, but behavioral and cognitive neuroscience research has failed to produce a common shared model. Facilitation is attributed to temporary lowered word recognition thresholds. Recent fMRI evidence identifies frontal<=>prefrontal<=>left temporal cortex interactions as the source of this priming bias.  The five experiments presented here clearly demonstrate that word recognition facilitation is a bias effect.  Context-Biased Fast Accurate Recognition (C-BFAR), a recurrent neural network model, shows how this anticipatory bias, proception, is accomplished by interactions between top-down conceptual-cognitive networks and bottom-up lexical word recognition networks.  Although classical signal detection theory says that the gain of facilitation bias is offset by the cost of miss-recognizing similar, but different, words, i.e., a zero-sum gain in accuracy, the prime typically creates a temporary time-space recognition window within which probability of the prime's reoccurrence is substantially raised – sometimes 1000-fold, paradoxically transforming "bias" into *de-facto* sensitivity. Given its facilitatory benefits, the word, bias, should be stripped of its pejorative connotation.

Recognizing each word and object in the continuous stream of our perceptual life is facilitated by our concurrent prehension of the multiple interacting contexts in which it appears. Reading one's email, listening to the news or walking down a crowded street, are all facilitated by neurocognitive expectation of the likely candidates of the next word or object in our environment. In the case of reading text, continuous updating of semantic and syntactic contexts enables our perceptual apparatus to anticipate many words so well that only a couple of letters may be needed for fast accurate lexical recognition. After reading this text you'll need only the letters *lex* to recognize the word *lexical*. Prior reading of a word, the prime, e.g.,, *lexical*, *biases* our neurocognitive word recognition networks so that on subsequent encounter of the same word as target, only a few of its letters are sufficient to accomplish accurate recognition.

Although the cognitive processes by which context facilitates word recognition have been studied extensively, the sharing of behavioral and cognitive neuroscience context procedures and facilitation models has been uneasy. One goal of this paper is to resolve a fundamental priming model assumption about which cognitive neuroscience and behavioral models have recently taken opposing positions. All neurocognitive priming models assume that facilitation produced by prior reading of the same word is accomplished by lowering the word's recognition threshold. (See, Collins, & Loftus, 1975; MacLeod & Masson, 2000; Neely, 1991; Taylor, 1953; Tweedy, Lapinski & Schvaneveldt, 1977; Van Berkum, Brown, Zwitserlood, Kooijman & Hagoort, 2005). In a major repetition priming research paper, however, Ratcliff and McKoon (1997) claimed that the facilitation of target recognition cannot be produced by any change in a "property of the representation of the word itself," such as lowering its recognition threshold. Their experiments were also precisely designed to determine whether repetition priming could produce a negative recognition effects when the primed word, the foil, was similar, but different from, the target word. The expected effect: inhibition of similar but not dissimilar foil words would help produce a more precise context recognition model. After describing this model, we show how a minor problem in their research procedure prevented them from getting their expected results and model. Correcting this problem, we carry out five experiments that resolve the conflict between cognitive and neuroscience priming models by demonstrating that recognition bias produced by repetition priming is indeed dependent on lowered recognition threshold of the primed word.

*Word Recognition By Bottom-Up-Top-Down Neural Nets Across the Cortex*

The major question addressed by every priming model is how does the prime lower the word's recognition threshold? Unlike other neurocognitive priming recognition models, C - BFAR assumes that the recognition threshold of a primed word is typically lowered by modest activation sent from the frontal cortex (fC) neural networks (NNs) that represented the original cognitive recognition of the prime words, via prefrontal cortex NNs, to the word's lexical NNs in the left medial temporal cortex (LTC) Lexical W layer (Figure 1).

Given the infinite size of PFC and FrC NNs, identifying their semantic and conceptual properties, interactions and values may be the most fascinating research challenge ever engaged by any science (See, Bielczyk, Llera, Buitelaar, Glennon, & Beckmann, 2019; Gazzaniga, 2004; Pauli, O'Reilly, Yarkonic & Tor, 2016; Rowe, 2019; and Yeo, 2015). Relative to lexical NNs in the LTC, information represented in PFC and FrC NNs is very abstract. Nevertheless, to understand the properties and function of those abstract NNs it will certainly be necessary to identify their relations to the lexical LTC NNs. Our increasing ability to identify neural representations in different cortical locations, particularly our improved access to FrC and PFC processes, suggests that there are infinite ways in which FrC context representations can facilitate recognition in LTC NNs. Perhaps *all* lexical word recognition is facilitated by FrC-LTC and FrC-parietal cortex interactions that lower the lexical recognition threshold of the expected word.

But this is just *part* of the facilitation process. Since the beginning of recognition neuroscience, word recognition models have been dominated by the *bottom-up* (BU) sequence of word recognition stages - from letter features in the occipital cortex, to letters, to lexical words, in the LTC. Because of the infinite complexity of context effects and the difficulty in identifying their many neurocortical sources, recognition experiments - methods and models –laboratory recognition procedures have traditionally been limited to one context at a time. Because they have not varied word context, they have been unable to isolate the unique contributions of context to the recognition process. So absent clear evidence of context on recognition, neurocognitive recognition models have been almost exclusively BU.

However, the BU synaptic pathways from the occipital cortex to LTC NNs are largely matched by interacting, *top-down* (TD) pathways. Some of the forebrain NNs are easy to identify because they represent cortical representations of "public" events such as words and objects. But the contribution of TD neural pathways to lexical word recognition is not limited to the forebrain. It is accomplished in BU⇔TD NN sequences running from the occipital cortex to FrC. The ability of PFC context to facilitate lexical word recognition in the LTC by lowering the lexical word threshold is, therefore, not a local lexical unit effect. As we show, context facilitation of lexical recognition is accomplished by interacting lexical and sublexical feature NNs that extend all the way back to the occipital cortex.

This paper does not model the construction of the interacting PFC⇔FrC context NNs. It simply shows how TD activation from these FrC context NNs facilitates lexical word recognition in the LTC. Three behavioral repetition priming experiments (N=274) identify the lexical facilitation effect produced by prior reading of the same word, and the recognition bias cost produced by prior reading of a similar, but different foil word. A lexical BU⇔TD NN model – from occipital cortex to LTC with TD input from the PFC – precisely simulates these behavioral facilitation and cost effects. They, in turn, fully support the Signal Detection Theory (SDT) assumption that the effect of PFC context on lexical recognition is one of *bias*. However, we emphasize that past objections to the STD facilitation bias *cost* assumption - that facilitation is cancelled by the cost of foil recognition, are a complete misapplication of SDT to real word reading environments.

We demonstrate these context recognition processes with a simple repetition priming procedure - the prior reading, minutes or hours earlier, of the *same word* - the *prime*. We demonstrate how these BU⇔TD interactions across 6 layers, from the occipital to LTC accomplish lexical recognition.

We show how TD PFC activation facilitates word recognition by biasing the lexical and sublexical networks.  For example, TD PFC context input that favors the *feet* in the Lexical W layer, lowers its threshold to BU input in several ways. It not only provides modest activation to the *feet* NNs, but modestly inhibits similar words but incorrect words, e.g., *meet*, in the Lexical W layer that share the *eet* subword letters.  And, TD activation from the  *f*  feature of *feet* inhibits the *m* feature in the L layer.  Thus, PFC context neurons lower the lexical word threshold by means of both TD and lateral activating and inhibiting processes, within each layer and between adjacent layers.

Unlike semantic and syntactic primes, the effect of repetition priming can last for hours, even days (Jacoby, 1983; Scarborough, Cortese, & Scarborough, 1977).  The processes by which the prime facilitates subsequent recognition of the same word as *target* should be easy to identify because they appear to be independent of any other context features or cognitive associations.  But despite extensive behavioral and neurocognitive repetition priming research (See, Bowers. 1999), the processes that produce context facilitation are not well understood.

Early models starting with the *logogen* word recognition model of Morton (1969, 1970, 1979; Murrell & Morton, 1974; Jacoby, 1983) proposed that priming temporarily lowers a word's recognition threshold. If so, how?  A plausible answer was that an episodic memory was created by the prime word's occurrence (Jacoby, 1983; Jacoby & Dallas, 1981; Masson & Freedman, 1990; Salaasoo, Shiffrin & Brooks, 1985).  To identify the characteristics of this prime memory, research employed a variety of recognition procedures such as requiring explicit judgments of whether the presented word was new, or old - i.e. previously presented  (See, Berry, Shanks, Speekenbrink and Henson, 2012; Cox & Shiffrin, 2017).  Frequent reports of conscious target recognition, i.e., "explicit" attempts to recall the prime, led Schacter (1990, 1992, 1994), Tulving, Schacter and Stark (1984), and Tulving and Schacter (1990) to propose that independent memory "systems" were required to account for explicit and implicit repetition priming.  A major debate asked whether repetition priming can be produced in the absence of conscious recall.  But extensive research demonstrated that repetition priming can indeed be produced entirely by implicit, non-deliberate, non-conscious processes in the absence of explicit-conscious target recognition processes (MacLeod et al., 2000; Masson and Bodner, 2003; Neely, 1991; Squire, 1992; Squire, Shimamura, & Graf, 1985; Squire & Zola-Morgan, 1991).  More recently, in a series of formal re-evaluations of the multiple systems controversy, Berry et al. (2012) have persuasively demonstrated that the evidence for multiple independent memory systems is not compelling.  They conclude that "a single memory strength signal drives recognition, priming and fluency is at least a viable alternative to the notion that there are functionally and stochastically independent explicit and implicit memory signals" (p 68).  But this still leaves the question of how reading the prime creates this temporary "memory," and how it facilitates subsequent target recognition, unanswered.

*Cortex Location of Word Recognition and Reciprocally Interacting Context Memory.* Substantial evidence points to the neocortex - prefrontal cortex, hippocampus and associated cortical regions anterior to the visual word form area, basal ganglia and the amygdala - as the source of several forms of transient memory of a prime (Buckner, Koutstaal, Schacter, & Rosen, 2000; Cer & O'Reilly, 2006; Norman & O'Reilly, 2003; O'Reilly, 2010). An early proposed contender for the neural source of repetition priming memory was the parahippocampus <=> hippocampus <=> cortex model in which a single recognition trial can establish a temporary link between cortical representations of the prime and its context features. It assumes that independent cortical representations of the prime word on the one hand, and of the priming context on the other - that have no preexisting link, can be temporarily linked via their prior independent pathways to the hippocampus.  Then, on subsequent presentation of the same word in the same context, the representation of the context via the recently established hippocampal connections can reactivate the cortical representation of the primed word - now the recognition target (See, Köhler, Crane, & Milner, 2002; O'Reilly, 2006, 2008; O'Reilly, & Rudy, 2001; Squire, 1992; Stern, Sherman, Kirchhoff, & Hasselmo, 2001).  Indeed, Cohen and O'Reilly (1996) proposed these prefrontal and hippocampal circuits as the basis of prospective memory - a memory that facilitates perception by anticipating future events. Repetition priming may be one component of this larger prospective memory system.  However, this model is incompatible with the finding that repetition priming survives among amnesic patients with profound hippocampal damage (Graf, Squire, & Mandler,1984; Steffanacci, Buffalo, Schmolck, & Squire, 2000; Berry, et al., 2012).

   Nevertheless, a damaged hippocampus may very well leave intact the large array of prefrontal cortex interacting networks that represent the cognitive and conceptual features of the lexical prime word, as well as the context in which it was recognized. We assume that the medial prefrontal "value" networks continuously act on these recognition networks to determine which should be converted to a permanent memory by the hippocampal circuits, which can be discarded, and which should be temporarily maintained as context for current recognition. We assume that the latter networks contribute to the "memory" that facilitates lexical recognition of a primed target.

   *Frontal and Prefrontal Lobe Contributions To Visual and Auditory Priming*. Given the greater convenience of visual over auditory word recognition and priming research procedures, lexical recognition has been more extensively studied and lexical neurocognitive recognition and priming models are more advanced than aural models.   Nevertheless, since in the learning life of every child, aural word recognition is acquired many years earlier than lexical word recognition, lexical neurocognitive structures and processes share many of the neurocognitive processes created by earlier aural word recognition. A functional MRI (fMRI) analysis of whole brain activity during repetition priming with visual and auditory words by Buckner, Koutstaal, Schacter & Rosen (2000) suggests how these shared PFC NNs may facilitate lexical recognition. They found reciprocal interactions between the LTC word recognition area and the PFC. The top-down leg of these interactions suggest that the fC and PFC circuits may be a/the basis for lexical priming facilitation.  The evidence, in brief, is as follows. Repetition of an unfamiliar word initially produces an *enhanced* fMRI response, i.e., an increase in blood oxygenation in the participating NNs. As the stimulus becomes familiar over several repetitions the magnitude of the fMRI response decreases. This distinction is confirmed by differential Gamma wave EEG responses to familiar versus unfamiliar words (Fiebach, Gruber, & Supp, 2005). In short, repetition priming of a familiar stimulus, such as a word, produces fMRI response "suppression" (a very misleading metaphor! Roediger & McDermott, 1993; Schacter & Buckner, 1998). We assume that this fMRI response decrement demonstrates that less sublexical input is needed to execute a second or third lexical recognition decision.

But while the cognitive conceptual features of lexical words as well as the features of the contexts in which they appear, are represented in the fC and PFC, there has been no clear evidence for how they facilitate lexical recognition. While examining the LTC, PFC and fC fMRI decrements to visual and auditory word repetition, Buckner et al. (2000) found that word repetition in *either* sensory modality produced the fMRI decrement in the posterior cortex recognition region (e.g., LTC for visual words) as well as in the forebrain regions that represent the higher order cognitive features of words. Unexpectedly, they found that word repetition in one modality also produced the fMRI decrement in the posterior cortex region of the *alternate* modality. For example, aural recognition of a word produced an fMRI decrement, not only in the forebrain, but also in the lexical LTC. Since the visual features of the word had not been produced by the visual word recognition - LTC, the fMRI decrement must have been produced by top-down input from the PFC. Although our research and model focusses on lexical priming, we will note where and how these processes interact and share those of aural word recognition and priming.

This suggests that during repetition priming the PFC and fC regions facilitate recognition processes in the LTC. This pattern of reduced cortical activity produced by repetition priming has been replicated and extended by Bergerbest, Ghahremani and Gabrieli (2004), Ghuman, Bar, Dobbins, and Schnyer (2008), Henson, Shallice, & Dolan (2000), Maccotta and Buckner (2004), and Reber, Gitelman, Parrish, and Mesulam (2005). Wig, et al. (2009) also showed that behavioral priming is accomplished by interactions among multiple PFC and fC lobes. Buckner (2010), Cole, Yarkoni, Repovs, Anticevic, & Braver (2012), and Siegel, Buschman & Miller (2015) found that that these locations are connected by "hubs" that maximize information flow across widely distributed cortical locations. The fronto<=>parietal cortex pathways appear to be the hub that supports repetition priming.

*How Does Top-Down Frontal to Prefrontal to Left Temporal Cortex Facilitate Lexical Recognition?* To put this question in perspective, we note that lexical word recognition is accomplished by a sequence of interacting BU⇔TD networks from the optic nerve and occipital cortex to the LTC. We assume that these BU⇔TD interactions continue through to the left fC where the cognitive conceptual and value features of each word is determined. At each intermediate step in this sequence, decisions are based on the joint input from BU and TD input. Note, that the information passed from pFC NNs to the representation of the lexical word is not a retrieved "memory," copied from one network to another, but rather cognitive and conceptual information previously associated with the lexical word in the fC NNs, passed down through the PFC hub pathway to partially activate the associated lexical representation in the LTC, or the Anterior Temporal Lobe hub, recently described by Hoffman, McClelland & Lambon Ralph, 2018.

We further assume that this fC TD input to the lexical word representation is sustained *only as long as appropriate* – for minutes or hours. (See, Collins & Loftus, 1975; (Collins, & Loftus, 1975; MacLeod & Masson, 2000; Neely, 1991; Taylor, 1953; Tweedy, Lapinski & Schvaneveldt, 1977; Van Berkum, Brown, Zwitserlood, Kooijman & Hagoort, 2005). And this context facilitation may be supported by the reader's interaction with her/his reading environment such as the computer screen and room location.

Given this sustained TD input, less matching BU sublexical input is required to complete the lexical word decision. This explains the fMRI repetition decrement in the LTC. In other words, lexical recognition is facilitated because less BU sub-lexical information is required to satisfy the lexical recognition threshold. Note, that the sustained character of this TD activation constitutes the reader's fC expectation of future repetition of the prime.

In this sense, the TD cognitive-conceptual associates of the word facilitate recognition of the primed word by temporarily *biasing* its lexical NN representation in favor of the recent prime word. The merit of a shift in bias rather than creation of a working memory is that bias does not alter the stable form of lexical word memory (unless, of course, it is a novel word not yet well learned). However, as we show below, the moderate activation of a primed lexical word representation by TD PFC input is a plausible candidate for the prime memory assumed by several priming models (See, Berry, et al., 2012; Cox et al., 2017; MacLeod et al., 2000; Masson et al., 2003; Rahnev, 2017; Sadeghi, McClelland, & Hoffman, 2014; and Schooler, Shiffrin & Raaijmakers, 2001).

*Context-Biased Fast Accurate Recognition (C-BFAR)*

Word recognition is carried out by a sequence of recurrently interacting NNs from the occipital to fC. Lexical word recognition, the traditional index of word recognition is but an intermediate step in re*cognition*. It is, however, the "public" evidence of word recognition because we can share it - we can record it. The final step, however, lies in the more abstract conceptual, cognitive representations located in the PFC and fC. Activated by prior input from lexical primes – fC NNs, via PFC NNs, continuously send activation back to the lexical word network in the LTC. In normal reading, lexical word recognition is accomplished by the reciprocal interaction of the TD cognitive and BU lexical recognition networks. By lowering the recognition thresholds of anticipated lexical input these PFC and fC NNs facilitate each successive moment of lexical, and perhaps all visual recognition. Context-based expectations are often so powerful that the exact word is sometimes "recognized" before it is presented (Masson et al., 1990). The strong participation of context in word recognition indicates a need to model the neurocognitive processes by which context information interacts with the BU word lexical word recognition process in order to facilitate recognition of the more abstract cognitive information signaled by the word (Berry, et al., 2012). Indeed, some computer-based text writing programs are quite successful in suggesting the next word that one is about to type.

In order to understand how priming facilitates word recognition it is necessary to have a clear model of lexical word recognition that is independent of priming. *Leabra* differs from the earlier connectionist models of McClelland and Rumelhart (1981) and Seidenberg and McClelland (1989) in that its learning processes are more compatible with biological models of synaptic learning and higher-order forebrain memory processes (O'Reilly & Munakata, 2000; O'Reilly, Munakata, Frank, Hazy, and Contributors, 2014). It replaces the back-propagation learning algorithm with recurrent pathways so that both BU and TD connections are learned between all adjacent layers, as well as lateral within-layer connections. C-BFAR shows how priming facilitates recognition of a primed word, how priming a similar but different word weakens target recognition, but priming a dissimilar word does not.

*Lexical Word Network Architecture.* The primary architecture consists of input letter Features (F), Letters (L) and lexical Words (W) layers (See Figure 1). These "public" layers, are interleaved with "hidden" layers that represent intermediate combinations of lower level information that enable interactions with the next higher level, and vice versa. Connections among all units are learned: between all units within each layer, e.g., between all words in the W layer, between all letters in the four possible letter units in the L layer, as well as among all possible letters within each letter position. Each unit in the layer between the L and W layers (Hidden LW) learns familiar combinations of letter strings in the Letters layer, such as the *ied*, that facilitate recognition of words such as *lied, died* and *tied*. Each letter pattern in the F layer is represented by a rectangular dot matrix (1s and 0s). Units in Hidden FL layer - the layer between the F and L layers may combine strings of these dots to represent simple line detectors such as the vertical and horizontal features of *l, t* and *f*. Consistent with cortical architecture, C-BFAR has recurrent connections within each layer, and between all units in adjacent layers, above or below. Unlike earlier connectionist models, *leabra* allows learning across overlapping input and output layers. In C-BFAR, both F and L layers may be treated as input layers, and L and W layer are treated as output layers so C-BFAR can recognize W from F, or from L, and its top-down connections also enable it to spell out the L that make up each W.

*Presentation of Degraded Target Word in Behavioral Procedure and C-BFAR Model.* As noted earlier, the degraded target features in the behavioral procedure are held in a temporary buffer followed by masking and reading the FC word cues. This buffer memory is fragile and may be modified by the recognition processes, but until we have studied it more thoroughly we cannot represent this process in C-BFAR. Therefore, C-BFAR leaves the degraded target in the input F layer with no descending connections from Hidden FL.

*Lexical Word Recognition Criterion.* In connectionist models such as *leabra*, word recognition is the process of competition among the representations of all words in the W layer for the best fit to the input. Given a word's bidirectional connections with the Hidden LW units, as well as with all other words in the W layer, especially similar, this process enables each word in the recurrent recognition cycles to improve the fit of the input pattern to a decreasing set of likely candidates. It continues until the word that provides the best fit to the data dominates all others. That is, further processing of the input produces decreasing improvement over time in the discrepancy between the input data and the criteria for the words in the network, i.e.

recognition error. This *settle* criterion, some asymptote of recognition error decrement over recognition cycles defines recognition accuracy and RT.  To model the forced-choice (FC) decisions of the behavioral experiments, competition is limited to the two cue words, continuing until one dominates the other.

*Recognition Bias: Lateral Inhibition is Greater Among Similar Than Dissimilar Lexical Words*. In the course of normal word learning, similar words, because of their similar BU input, especially from subword letter set in C-BFAR Layer LW, acquire greater mutual (lateral) inhibition in the W layer than do dissimilar words, which because they have dissimilar BU input do not compete. Lexical word competition is greatest among words that share a similar subword letter sets, e.g., *foot, loot, moot, coot* and *soot* (See, Hidden LW, Fig. 1).  BU activation from any *one* of the five words sends identical *oot* activation to all five similar words in the Word layer. Accurate recognition of the target, e.g., *foot*, can be accomplished only when the distinctive letter, *f…* , provides *foot* with an increment in activation sufficient to suppress competing activation from the similar, but false words,  *loot, moot, coot*  and *soot*.

As C-BFAR demonstrates, this greater lateral inhibition among similar words is *the* basis of the repetition priming bias effect of among similar words identified by Ratcliff and McKoon (1997). The experiments described below replicate that priming bias effect.  Note, that this greater lateral inhibition is why the degraded target words require longer exposure time in the Similar 2FC experiments to reach the same baseline accuracy as targets in the Dissimilar 2FC experiments.  The Ratcliff and McKoon failure to find positive priming in the Dissimilar 2FC condition was because they used the same target exposure time for Similar and Dissimilar 2FC decisions. The lower lateral inhibition in the Dissimilar condition produced a higher baseline accuracy and a ceiling effect that was insensitive to target prime facilitation.

Note, that inasmuch as SDT assumes that signal and foil are independent of these interacting sub-lexical processes it cannot accurately represent bias produced by priming one (e.g., *died*) versus the other (e.g., *lied*) as a criterion shift in favor of the primed word.

*Learning the Word Set.*  C-BFAR employs 150 4 letter (consonant (C), vowel (V),C,C,) words each of which have at least one similar word, i.e., that shares a common three-letter string. As described earlier (Fig. 1), the connections between, and within all layers are learned simultaneously. Recognition strength of a word (W) is indexed by its activation level in the W layer (W ACT). Accurate recognition occurs when a W's ACT settles at its asymptote, typically 0.87, at which point the ACT of all other Ws drops < .02.  RT is indexed by number of cycles to reach this criterion.  The entire 150 word set reaches this criterion in fewer than 50 learning trials.

*Recognition in the FC Context*.  As in the behavioral procedures, where recognition accuracy is the dependent variable and the target word recognition decision space is confined to 2 words.  Target word Fs are severely degraded and masked, and cannot be recognized without the assistance of the two FC cues.  To match the behavioral Similar and Dissimilar baseline accuracy means in Exps. 2-5 of 0.75, Fs are masked more severely in the Dissimilar than Similar condition. C-BFAR represents the contribution of the FC cue of the behavioral recognition procedure as modest TD activation (FC ACT) to the lexical representations of the target and foil FC words, which lowers their recognition thresholds. C-BFAR demonstrates how FC ACT interacts with the BU lexical recognition network to enhance the interactions that are most consistent with one of the two FC Ws and suppress less consistent alternatives. Recognition of a target word is initiated by activating its masked F input in the recognition network of 150 words while simultaneously sending FC ACT down to *both* target and foil representations in the W layer. The interacting BU⇔TD recognition competition stops when one word dominates the other in the W layer. Baseline accuracy = number of correct target words/150.

   Initial C-BFAR simulations suggested that all words were not participating equally in this process, so we examined the distribution of target W ACT in the absence of TD FC ACT. W ACT distributions were extreme - many words quite high while many more were = 0. Therefore, all subsequent simulations were computed after a tedious process of setting the mask strength individually for each word, separately for Similar and Dissimilar treatments, much as we had set behavioral baseline accuracies at .75 by adjusting target exposure durations for each reader. Wolfgang Pauli designed a leabra program, *fit*, that allows one to select values of the FC, prime, target masks and other parameters of C-BFAR. To avoid recognition values being the artifact of a particular *pattern* of mask features, all baseline and prime means were based on multiple simulations, each with different mask patterns.

   *Priming*. C-BFAR assumes that primed PFC and Fc NNs send modest activation back down to their source, the lexical Ws. To simulate this top-down effect, C-BFAR sends modest activation to the primed FC lexical W. A small Prime ACT of 0.019 interacts with the BU recognition NNs to produced identical $P_{TS}$ and $P_{TD}$ primes of .05, matching the $P_{TD}$ of Exps. 2 + 3, and the $P_{TS}$ of Exp 5. That the pattern of the four C-BFAR primes match the behavioral primes supports that assumption that the W layer of the lexical Word recognition network is the place where TD PFC and fC recognition contexts act on lexical word recognition.

   In the course of simulating the behavioral prime values, C-BFAR found that as mask severity increased, the .75 Similar Baseline value could be maintained simply by raising FC ACT. And the Similar target prime, $P_{TS}$ of .05 could be generated with different combinations of mask severity and FC ACT as long as Similar Baseline = .75. However, the behavioral $P_{FS}$ of -.04 could be obtained *only* when FC ACT = 0.084. For FC ACT values > .09, the absolute value of the negative $P_{FS}$ prime became equal or larger than the $P_{TS}$ of .05. Why? Consider how Foil ACT produces $P_{FS}$. The Hidden LW layer plays a major role in recognizing a Similar word. The letter strings, such as the *ood* of *food, good, hood, mood* and *wood*, are represented by a single learned unit in the Hidden LW layer. And when any single target word in that set is read, e.g., *food,* the *ood* unit sends activation to every *ood* word in the Words layer. The single letter*, f,* that distinguishes the correct target, *food*, from all other words would have little effect if its contribution were merely additive. Fortunately, in the course of word learning, the *leabra* neural network model learns to compensate for this confusion by strongly inhibiting all the similar, but incorrect, words. Supporting that assumption, it is this lateral inhibition learned as part of word recognition that accounts for the large negative Similar foil prime, $P_{FS}$. Having little inter-word competition Dissimilar words need little or no lateral inhibition, have a near 0 foil prime.

*Priming Bias Facilitation minus Cost.*  The cost of this bias, of course, is a corresponding increase in recognition error - misrecognition of similar, but different (foil) words. Since this higher foil threshold increases the probability of *not* recognizing the foil, Signal Detection Theory argues that the net effect of priming for 2FC decisions is zero – which it is if target and foil are equiprobable. In reading a book or newspaper article, however, where there may be many similar foils, e.g., *face, fake, fare, fame,* the recognition cost could far exceed the facilitated recognition of the primed word.  As we will show, however, the prime usually introduces a recognition environment context in which the probability of future repetitions of the prime word are greatly increased, sometimes 1000-fold, while the probability of similar foil words remains unchanged, near 0. Therefore, there is *no increase in recognition cost*, and the net effect of priming is facilitation.  Biasing the recognition system in favor of the primed word, saves recognition time, neural processing resources, with little or any loss in recognition accuracy.

*Repetition Priming Bias:  Behavioral Evidence*

The probability that repetition priming biases target recognition was initially proposed and demonstrated in a series of experiments by Johnson & Hale (1984), Ratcliff, McKoon & Verwoerd (1989) and Ratcliff et al., 1997).  As noted earlier, a severe ceiling effect in one experiment produced results that were incompatible with all previous priming models and its model could not be mapped onto word features.  After reviewing this experiment, we describe five experiments that reexamine and fully support the repetition priming bias assumption.  C-BFAR has shown how these behavioral findings support a multi-layered NNs from the visual line features in the occipital lobe, to lexical word recognition in the left frontal lobe, and the top-down input from pre-frontal and frontal lobes accurately simulates the pattern of repetition priming obtained in the behavior experiments.

Building on earlier SDT studies (Johnson & Hale, 1984; Ratcliff et al., 1989; Ratcliff et al., 1997, Exp. 1) designed a sensitive target recognition procedure where each degraded target is followed by two FC alternatives – target (correct) or foil (incorrect).  They measured target recognition accuracy under three conditions: target primed, foil primed, or no prime (i.e. baseline), which yield two primes: target (target minus baseline accuracy) and foil (foil minus baseline).  Crossed with two conditions: Similar FC alternatives, e.g., *died/lied*; and Dissimilar, e.g., *died/sofa,* these procedures produce the four primes needed to identify priming bias: Similar FC target prime ($P_{TS}$), and foil prime ($P_{FS}$), and Dissimilar target ($P_{TS}$) and foil prime ($P_{FS}$). SDT suggests that priming the target should produce positive $P_{TS}$, and $P_{TD}$ primes.  Priming the foil should produce a negative $P_{FS}$, but have no effect on $P_{FD}$.

*Target Recognition: A Closer Look.*  To ensure that target recognition engages the full range of features that distinguish target and foil words, the traditional FC recognition sequence required modification.  The traditional recognition sequence -  FC cue => target => mask => FC response –where FC cues precede the target, encourages the reader to optimize recognition accuracy in both the Similar and Dissimilar FC decisions simply by limiting attention to one informative letter position, e.g., ***died/lied*** in the Similar, and any letter position, e.g., ***died/sofa***, in the Dissimilar. But since priming bias acts on whole words this recognition strategy would make the procedure insensitive to the difference between Similar and Dissimilar FC decisions. Ratcliff and McKoon solved this problem by placing the FC cues *after* the target: - target => mask => FC cues => FC response. But this sequence creates a new problem. The degraded target features must be "saved" in a temporary memory location while the same BU word recognition network is employed to recognize the FC words. Is the memory of these target features further degraded by the recognition process itself? The following experiments address this question.

*Repetition Priming with Dissimilar FC Target and Foil, and the Ceiling Problem.* For recognition accuracy to be equally sensitive to both positive ($P_{TS}$ and $P_{TD}$) and negative ($P_{FS}$) target primes, baseline accuracy needs to be close to .75 in both Similar and Dissimilar FC conditions. Accuracy distributions tend to be increasingly negatively skewed as accuracy increases, which makes them increasingly insensitive to positive bias. In two critical experiments with the modified FC recognition sequence, Ratcliff and McKoon found qualified support for the bias assumption. When target and foil were similar, priming the target significantly increased target accuracy and decreased response time (RT); priming the foil decreased target accuracy and increased RT (1997, Table 1; N=16 + 16). However, they found no evidence of bias when the FC alternatives were dissimilar, i.e., $P_{TD} = 0$. Therefore, they were obliged to dismiss all models that assume that "prior exposure to a word changes some property of the representation of the word itself. For example, the resting level of activation of the word…" p 339. They proposed a counter model - a diffusion model - derived from the random walk model in physics, in which the FC target and foil each has a decision counter that accumulates evidence supporting a recognition decision. Prior exposure to a word causes its counter to become an attractor enabling it to steal counts away from the counters of other similar words, in this case the unprimed Similar FC alternative (Izhikevich, 2010). Because the priming effect is weak and Dissimilar FC alternatives share no features, priming the Dissimilar foil does not enable it to steal counts away from the target alternative. A major limitation of this counter model is that its functions cannot be mapped onto any properties of word, or sub-word features so that the recognition decision is made outside of the recognition networks. Furthermore, since a counter has no features it cannot be an attractor.

Bowers (1999) pointed out that a severe ceiling effect obscured evidence for a positive Dissimilar FC target prime effect in the critical Ratcliff et al. (1997) Dissimilar FC experiment. Relative to the Similar FC decision, e.g., *died / lied*, the Dissimilar decision e.g., died / sofa, requires fewer target features to make a correct recognition decision so that when target degrading was held constant across the two FC conditions, baseline accuracy was substantially higher in the Dissimilar. While Ratcliff and McKoon obtained a baseline accuracy of .68 in the Similar FC condition, it increased to .84 in the Dissimilar baseline. Indeed, the fastest 2/3rds of the participants in one Dissimilar condition had a Baseline accuracy of .93, suggesting that some subjects had a Dissimilar Baseline of 1.0. The negative skew of accuracy data and the high baseline mean made the Dissimilar condition much less sensitive than the Similar FC procedure to a positive target prime effect. This conclusion is confirmed by Exp. 1 (N = 73) of this paper.

In a further review of Exp. 4 (Ratcliff et al., 1997) and Ratcliff, et al. (1989), Bowers (1999) identified several experiments where, when ceiling effects were minimized, the Dissimilar FC target prime magnitude ($P_{TD}$) were close in magnitude to the Similar FC target prime magnitude ($P_{TS}$). To demonstrate floor and ceiling effects on prime magnitudes, Bowers then carried out several experiments using the Ratcliff et al. (1997) procedure - except that he varied the duration of target presentations between Similar and Dissimilar conditions in order to achieve different levels of baseline accuracy. Baseline accuracy ranged from near ceiling (with flash time = 25 ms) to near chance (flash time = 9 ms). The predicted ceiling suppression of $P_{TD}$ magnitude was confirmed. In sum, although Ratcliff and McKoon's null $P_{TD}$ evidence led them to reject all repetition priming threshold models, Bowers' (1999) experiments discount their basis for that rejection. We now present five experiments and a neural network model that confirm the assumption that the facilitating effect of repetition priming is produced by biasing the lexical word recognition process.

### Five Repetition Priming Experiments

*Summary of Experiments.* The first goal of this paper is to obtain accurate estimates of the four prime magnitudes ($P_{TS}$, $P_{FS}$, $P_{TD}$, $P_{FD}$) by using large samples and matching baseline means and accuracy ranges in the Similar and Dissimilar FC procedures so that any floor or ceiling constraints are identical in the two procedures. The primary independent variables are those of Ratcliff et al. (1997) as described here in Table 1: Exp. 1 uses both Similar and Dissimilar FC trials, 2, 3 and 4, use only Dissimilar, and 5, only Similar. Response times (RTs) from FC onset to response were also recorded. For Similar FC Exp. 5, N=105, baseline accuracy, BS = .75; for Dissimilar Exp. 2 & 3, N=169, baseline accuracy, BD = 0.75. Across subjects, within treatment conditions, baseline accuracy values are negatively skewed and can vary widely so that some participants could have near perfect baseline accuracies that are insensitive to a positive priming effect. To avoid these individual ceiling effects, we pre-tested each reader and restricted individual pre-experiment baseline target flash times so that each reader's accuracy fell within the .65-.85 interval. The prime to target recognition interval was approximately 15 min. To obtain accurate estimates of prime magnitudes we substantially increased the number of trials per condition and obtained a combined participant sample size over three experiments, of 274. We find identical Similar and Dissimilar target primes ($P_{TS} = P_{TD} = .05$), a Similar foil prime that was slightly smaller than the target prime ($P_{FS} = -.04$ ), and a Dissimilar foil prime ($P_{FS}$) of 0.

*Word Stimuli.* Prime, target and *FC* words were selected from lists of triplets, constructed after the procedure described by Ratcliff et al. (1989). Except in Exp. 3, word triples, (N = 82) were constructed so that two words in each triple were similar to each other in that they differed by only one letter, which appeared at any position of the word, and the third was dissimilar so that not only the letter, but the letter features, differed in each letter position. Exp. 3 used quadruples of 3 similar and 1 dissimilar word. Word length ranged from 3 to 7 letters. Mean word frequency = 140/million; mean log frequency of use in English (Carrol, Davies &

Richman, 1971) was equivalent across all words in the triplet or quadruplet. All words were presented in lower case.

*Priming Procedure.* We reduce the use of explicit retrieval processes at recognition by slightly modifying the priming instructions and procedure of Ratcliff & McKoon. Rather than instructing the readers to learn the prime words for "a later (unspecified) test" (p 322, Ratcliff et al., 1997), they were asked simply to check the words for spelling errors (10% of all words), indicating their judgment on a toggle switch. The spelling decision forces readers to process all of a word's letters thereby avoiding the more common reliance on the first letter of a word (See, Scaltritti & Balota, 2013). Primes were presented on the screen for 1sec., followed by a mask followed by a pair of asterisks (**).

*Recognition Test: Target Recognition Accuracy by Duration Calibration.* Target stimuli subtended approximately 5º of an arc at the reader's eye. Except in Exp. 1, a training and calibration procedure was used to estimate the target recognition accuracy by duration for each participant in the test phase of the experiment. The recognition task was initiated by pressing the space bar. After 'Get Ready' appeared in the center of the screen, the display remained blank for 1000ms; the target was flashed on the center of the display, followed by the mask, a row of '@@@' signs, that completely covered the target area. Mask duration was 500ms, except in Exp. 1 where mask duration was a treatment variable. The mask was followed by a side by side presentation of the *FC* words for 2000 ms, or until the participants made a response, whichever came first. Participants decided whether target best matched the left or right *FC* word, and pressed "1" or "2" key on the number pad to record their decision. Participants were asked to make their decision quickly but also to be as accurate as possible.

Target durations were of fixed duration in Exp. 1; but in all other experiments, an accuracy-duration function was computed for each reader to achieve a sample mean baseline accuracy, $B_{Acc}$, of 0.75, and a .55-.90 range. Unless otherwise noted, individual participant exposure durations were determined in 40 pre-experiment calibration trials. Following 5 trials @ 100 ms to learn the recognition procedure, 40 duration calibration trials were presented, dropping from 35 ms. to 15 ms. in 5 ms. steps, 8 trials per step.

*Equipment.* Stimuli were presented on 17" PC monitors. Presentation of the words and recording of correct and error RTs were computed by a custom-built program.

*Participants.* All participants were volunteer undergraduates. In a pilot study we found that an N of 31 provided too little power to identify prime magnitudes as small as .05, so we attempted to run all experiments with N > 80 to assure a power of .8 for $\Theta^2 = 0.1$. English was the primary language of all readers; their vision was normal or corrected.

Experiment 1

Assuming a severe ceiling effect might eliminate a Dissimilar target priming effect, this experiment was designed to replicate Bowers (1999) finding that there are severe ceiling and floor accuracy effects when the same target exposure duration is used for Similar and Dissimilar FC recognition. It also asked whether there is an FC order effect, such that accuracy is lower for left than right FC words, and if such an effect interacts with mask duration. If it does, the difference should be larger for a short (300 ms.) than long (1,500 ms.) mask.

*Method*

*Prime Set*. The set consisted of 12 primes for each of Similar and Dissimilar targets and foils, (TS, FS, TD, and FD) trials, plus 12 misspelled words.

*Design: Study and Recognition Test.* The experiment is a mixed design: between: 2 mask durations by within: prime condition (target primed, foil primed, or no prime word), Similar or Dissimilar FC cues (See Table 1), presentation order (target presented on the left or the right), and 73 participants. Similar and Dissimilar FC trials were blocked and their order counterbalanced across participants.

*Target Duration and Mask.* Target durations for the first 58 subjects were 30 ms. across both Similar and Dissimilar conditions so that *letter* features, would be constant across treatments. Targets were followed by a brief mask. However, 30 ms. yielded few participants in the lower accuracy range in the dissimilar FC trials so target flash time was reduced to 20 ms. for the last 15 subjects.

*Participants*. Of the 75 volunteers in this experiment, two were dropped because low accuracy on 3 or more conditions, making N= 73.

*Results*

*Mask Duration*. Mask duration had no effect on baseline accuracy in the first 58 subjects ($F_{1,56}$ <1.0) so data for the two duration conditions were collapsed for subsequent analyses and a mask duration 500 ms. was employed for the final 15 subjects.

*Accuracy and Primes: Similar and Dissimilar Trials*. Mean Baseline Accuracies: Similar BS = .65, Dissimilar BD = .80, Mean =.73. Similar primes, $P_{TS}$ and $P_{FS}$ were .07 and -.01 respectively; Dissimilar primes, $P_{TD}$ and $P_{FD}$ were .03 and -07.

*Prime Magnitude across the Baseline Accuracy Range.* All four prime magnitudes were subject to the ceiling effect. Across subjects, all four prime magnitudes - $P_{TS}$, $P_{FS}$, $P_{TD}$, $P_{FD}$, were a negative function of baseline accuracy with slopes ranging from -.32 to -.59 . The largest absolute primes, $P_{TS}$ and $P_{FD}$, were those least vulnerable to floor or ceiling effects. With BS = .65, the positive $P_{TS}$ = .07 (p<.001), while with BD = .80, the negative prime, $P_{FD\_Acc}$ = -.07.

*Response Times.* Mean baseline RT = 749 ms; BS RT was 84 ms slower than BD RT ($F_{1,68}$ = 66.7, p <.001). Target RTs, $P_{TS}$ and $P_{TD}$ , were 40 ms faster and Foil RTs, $P_{FS}$ and $P_{FD}$ were 20.5 ms slower than their respective Baseline RTs ($F_{1,68}$ = 6.0, p <.02, $F_{1,68}$ = 3.8, p <.06, respectively).

*Discussion*

*Similar Versus Dissimilar FC Target Recognition Decisions.* The faster and more accurate Dissimilar Baseline values demonstrates the smaller magnitude of target feature information required for Dissimilar than Similar FC recognition.

*The Effects of Priming on Recognition Accuracy and Speed*. Priming the target increased both accuracy and speed, while priming a similar but different word decreased speed and accuracy. Primes that had the smallest absolute magnitudes were those most vulnerable to floor or ceiling effects, $P_{FS}$ and $P_{TD}$, confirming Bower's conclusion that estimated prime values are vulnerable ceiling effects associated with variation in baseline accuracy. Consequently, subsequent experiments established both Similar and Dissimilar mean Baseline values at 0.75.

Experiment 2

This experiment is tests whether a Dissimilar FC target priming effect occurs when target durations are set so that BD = 0.75. It further determines whether priming effects across participants are a function of individual baseline accuracies.

*Method*

*Participants.* Seventy-five students were paid for their participation.

*Pretest calibration of Target Durations.* Target durations were calibrated for individual participants to target a mean BD = .75. No durations exceeded 35 ms. Mask duration was 500 msec. To improve the power number of baseline trials was tripled to 36. To avoid subject fatigue, FC alternatives were limited to the Dissimilar condition.

*Stimuli.* The word triples for pretest, study and test phases of Exp. 1 were modified to use only matched dissimilar FC words. In the recognition procedure, order of the FC test words was counterbalanced in blocks of 4 across participants.

*Results*

*Recognition Accuracy and Priming.* BD = .75; RT = 626 ms. $P_{TD}$ =.05 (Table 2; $F_{1,75}$=10.12, p=.002). $P_{TD}$ was equivalent in both halves of the session, thereby demonstrating that the effect of the prime did not deteriorate across trials or time within the session. There is a small $P_{FD}$ of -.02 of borderline significance ($F_{1,75}$ = 3.68, p = .06 ).

*Regression of Prime Accuracy onto Baseline.* As in Exp. 1, we computed the regression of $_cP_{TD}$ and $_cP_{FD}$ onto BD after the RegTTS and FC order corrections (based on a 2 by 2736 - 76 participants X 36 trials - data set). Without the corrections, the regression of $P_{TD}$ on BD was substantial and significant ($R^2$ = .11, $F_{1,74}$ = 9.09, p =.004). After correction, neither were ($_cP_{TD}$ on $B_{UD}$, $F_{1,74}$ = .12, p=.73).

*Discussion*

The significant Dissimilar FC target prime, $P_{TD}$ of 0.05 when BD = 0.76 supports Bower's assumption that Ratcliff & McKoon's failure to find a Dissimilar FC target prime was due to a ceiling effect, and suggests that priming is not affected by the distance in between alternative counters in the Ratcliff & McKoon model. The regressions of prime magnitudes across baseline, corrected for regression toward individual subject baseline means, indicates that in the present data set, prime magnitude is independent of floor and ceiling constraint.

Experiment 3

Bowers (1999, 2000), Wagenmakers, Zeelenberg, Schooler, & Raaijmakers (2000a & b), and Zeelenberg, Wagenmakers, & Raaijmakers (2002) have reported significant positive $P_{TD}$ values. McKoon &Ratcliff (1996, 2001) and Ratcliff et al. (1997) attributed those findings to explicit recall of the prime words at test. They suggested that Bower's casual prime instructions - to simply read the primes rather than the Ratcliff and McKoon memorize the prime words, did not create a strong prime memory for the primes so that subjects were obliged to use a more explicit prime recall procedure at test, which in turn produced his positive dissimilar target prime. McKoon and Ratcliff (2001) compared the two instruction procedures and used a debriefing procedure to obtain subjects' subjective reports of whether they thought they had used explicit recall of study i.e., prime, words at test. They found that while "none of the participants receiving (the) ... memorization instructions (N = 14) said that they had intentionally chosen words from the study lists on the *FC* tests, all of (those) receiving Bowers' (1999) instructions (N = 14) either volunteered or agreed that they had sometimes done so" (p 676). Their data strongly imply that Bowers' more casual priming instructions -- "Read these words..." produced an accuracy prime based on an explicit retrieval of prime information. Inasmuch as their criticism might apply to our findings we compare the effects of our Spelling and their Memorize instructions on $P_{TD}$ magnitudes. In so doing we note that our Spelling instructions are *more* implicit that the Ratcliff & McKoon instructions "to learn the (prime) words for a later (unspecified) test."

## Method

*Design*. The experiment was a mixed design: 2 instructions (Priming instructions: Memory vs. Spelling) by 2 prime-test set sizes (2 vs. 4 sets) between-observer factors, by 3 classes of FC decisions (UD, TD, FD) – fully crossed, with 2 random orders of the study words in each condition combination. The 110 participants, most of whom were pre-med students, were randomly assigned to the 2 study conditions.

*Distribution of Target Exposures*. To provide a more sensitive test for possible floor and ceiling priming magnitudes within the sample a larger range of individual pretest accuracy values was permitted than in Exp. 2, while targeting baseline accuracy at .75.

*Stimuli*. Words for all conditions were selected from a new set of similar triples which differed by only one letter, matched to one dissimilar word that had the same number of letters or letter shapes, matched as closely as possible in word frequency. Targets were 18 each of target, foil and baseline words.

*Study Instructions*. The Memorize condition used the Ratcliff et al. (1997) "learn" instructions; the Spelling condition, described above, asked participants to check the words for spelling errors, where 10% of the primes were misspelled.

## Results

The initial N of 110 produced Spelling and Memory baseline means of .75 and .73 respectively. Examining the magnitude of $P_{TD}$ and $P_{FD}$ as a function of BD and BD-Median RT we noted 3 participants who exhibited the putative response pattern of explicit priming: modest BD ( <0.6), large primes (both |$P_{TD}$ and  $P_{FD}$ | >0.25), and slow  BD-Median RT (>1200 ms). All were from the Memory Study condition. Dropping them from the sample was a conservative decision in that dropping these participants eliminated the cases with the highest absolute magnitudes of $P_{TD}$ and $P_{FD}$. We were also concerned that some participants might not have understood the target recognition instructions, or were not well practiced. Examination of the data located 4 participants who had missed >.10 trials. We therefore dropped 9 participants, leaving a total N of 101: 49 in the Spelling and 52 in the Memorize conditions respectively. Ns for the two and four study sets were 24 and 25 for the Spelling and 27 and 25 in the Memorize conditions respectively. The standard deviation of BD increased to .17, up from .14 in Exp. 2.

There were no main significant priming or *RT* effects for Set Size (all $F_{1,100} < 1$, n.s.).

*Primes.* The significant $P_{TD}$ of .05 (Table 2; $F_{1,100} = 27.9$, $p < .01$ ) replicates that of Exp. 2 (See, Table 2). $P_{FD} = .02$ ( $F_{1,100} = 3.61$, $p = .06$ ).

______________________________

Place Table 2 and Figure 4 about here

*Study Instructions, Number of Prime-Test Sets, and Priming.* Comparison of Spelling versus Memorize primes showed no differential priming magnitudes. Spelling and Memorize $P_{TD} = .047$ and .048 respectively. Neither $P_{TD}$ nor $P_{FD}$ RTs interacted significantly with the Spelling-Memorize treatments ($F_{1,99} < .10$, n.s.).

*Target Decay By FC Cues.* Given that the degraded target must be held in some location other than the lexical network while that network reads the FC words, one must ask whether that target information is eroded by reading the FC words and further, by the target recognition process that attempt to match the target to one of the FC alternatives. Assuming a Left-Right reading order, BD was .09 lower for Left, relative to Right FC cues. Given the lower Right BD of .70, the Right prime was significantly larger than the Left (.08 versus .02).

## Discussion

*Dissimilar Target Prime.* The mean $P_{TD}$ for the two experiments of .05 (N = 169) strongly supports Morton's logogen model and all models that assume repetition priming lowers the recognition threshold of the primed word. The not quite significant, but opposite $P_{FD}$ values (-.02 & +.02) suggests that $P_{FD}$ is close to 0.

*Explicit Episodic Bias.* The almost identical Memorize and Spelling instruction primes provides no support for the assumption (McKoon et al., 2001) that this substantial Dissimilar $P_{TD}$ can be attributed to explicit retrieval bias. That the three participants who appeared to be using an explicit retrieval process at test all came from the Ratcliff & McKoon Memorize instruction condition argues against the position that the weaker "Spelling" instructions increase vulnerability to explicit recall at test.

## Experiment 4

Indeed, when degraded target information is close to 0, readers may select the primed FC word simply because it is the more familiar of the alternatives. Bowers (1999) carried out such a test using non-letter symbols such as $*&#@ flashed at 9 ms., and found positive target and negative foil primes in both similar and dissimilar conditions. Concerned that the observed effect might be "partly contaminated by explicit memory strategies" (p 590) he repeated the experiment with a deadline procedure designed to restrict attempts at test to recall words in the prime set and found somewhat diminished effect sizes. On the strength of his findings, and the assumption that our Spelling instructions -- "Check these words for spelling errors." --- were much less explicit than the Memorize instructions of Ratcliff and McKoon, we further examine the functional role of explicit processing in priming by estimating cross-participant correlations between performance and subjective estimates of prime memory strength and the use of explicit

processes. The primary goal of this experiment, however, is to determine whether priming does, or does not bias FC decisions when the target possesses no information.

## Method

*Design and Stimuli*. Fifty one participants were randomly assigned to one of two priming study instruction conditions: the Spelling instruction employed in all prior experiments here, or the Ratcliff and McKoon Memorize condition. Study lists contained 36 words, plus misspelled words in the "Spelling" condition. There were four FC target conditions: one with alpha-numeric non-word targets (Bowers, 1999) where one of the *FC* pair was studied, and the three standard test conditions where the degraded target word had either been studied, its foil studied, or neither. Individual target durations were set so that mean BD would be close to 0.6. There were 10 practice and 90 test trials in this phase. The experiment was a mixed design with 2 between-factor levels (Memory or Spell study instructions) and 3 within-Subject levels of study conditions (target, foil, neither target nor foil studied) crossed with word versus non-word targets.

Non-word targets were constructed with the following constraints: They had the same numbers of letters and vowels as the words of the *FC* pair at test, and the letter in each position shared as few as possible linear or curvilinear features with the letters in that position in either *FC* word. For example, for the FC pair, [deep/soot], the target might be [ muki ]. The random letter targets were flashed for 5 ms. Pilot runs revealed that some subjects reported that real words were *never* being presented and therefore stopped trying to recognize the targets. To mitigate this problem, non-word targets were interleaved with degraded word targets.

*Debriefing*. To assess the prime memory magnitude at the conclusion of the recognition procedure, participants read a list of words - the full set of primes mixed with an equal number of non-prime words, and checked those that they recognized from the prime procedure. To assess their confidence that they had used an explicit recall process at test, they reported whether they had ever thought, *while responding to the targets*, that either of the FC alternatives were the same as the prime words, and if so, on what percentage of the trials.

## Results

*Baseline Accuracy: Words*. $BD_{MEAN} = .58$; $BD_{SPELL} = .61$, $BD_{MEMORIZE} = .54$ .

*Priming Non-Word*. When targets consisted of nonwords, so that $BD = .50$, $P_{TD}$, n probability of choosing the primed *FC* alternative was 0.0 ( -.06 for the Spelling condition ($F_{1, 25} = 4.2$, $p = .05$), +.06 for Memorize ($F_{1, 20} = 4.8$, $p = .06$).

*Prime Magnitude and Subjective Estimate of Explicit Processing*. The proportion of trials on which participants estimated that they engaged in explicit recall at test was *independent* of $P_{TD}$ ($r=.06$, n.s.), and the proportion was identical in the Memorize (34%) and Spelling (36%) conditions. The relation between subjective prime strength and priming magnitude was modest.

## Discussion

The assumption that when target information approaches 0, priming biases FC decisions in favor of the primed FC alternative is not supported. Furthermore, the McKoon et al. (2001) finding that weaker prime instructions induced explicit recall at test was not supported. Readers' estimates of using explicit recall at the point of target recognition was independent of whether priming instructions employed the more explicit study emphasis of the Memorize instructions or the weaker implicit Spelling instructions. Therefore, the positive Dissimilar target prime in Exps. 2 & 3 is independent of explicit memory processes at the point of target recognition. The mean $P_{TD}$ of .05 is the result of a decreased target word recognition threshold which we attribute to an increase in the activation of prime word's lexical representation.

## Experiment 5

This experiment addresses two questions. The diffusion prime model says that the decrement in target accuracy produced by priming the similar foil is equal in absolute magnitude to the positive increment produced by priming the target itself, i.e., $|P_{FS}| = P_{TS}$. But as we shall see, C-BFAR suggests that the indirect effect of a foil prime on target representation, $P_{FS}$, may be smaller than the direct effect of the same prime target, $P_{TS}$. Although Exps. 2 & 3 obtain significant Dissimilar primes, their magnitude of $P_{TS} = .05$ is substantially smaller than the Similar primes of .10 - .12 reported by Ratcliff et al. (1997). By employing the same Spelling instructions and single study-test format as our prior experiments, we can determine whether any difference in $P_{TS}$ is due to differences in experimental procedures, or to true differences in $P_{TS}$ versus $P_{FS}$ magnitudes.

Prior studies have been unable to address these questions because they have not matched Similar and Dissimilar baseline accuracies and minimized baseline ceiling and floor limits. Using the same individual pretesting procedures as the prior experiment this one sets $B_{Acc} = .75$. Because more letter features are needed to reach a Similar, e.g., *died/lied*, than Dissimilar, e.g., *died/sofa*, $B_{Acc} = .75$, Similar targets must have longer target exposure times. A Similar FC decision requires feature information from nearly all target letters, whereas a Dissimilar target can be recognized by partial features of a single target letter. Nevertheless, by setting both Similar and Dissimilar FC $B_{Acc} = .75$, target information is identical in the two conditions.

*Method*

*Design and Word Sets.* Stimuli consisted of the same 72 word triplets as in Exps. 1 and 2, but in this experiment *only* the Similar FC pairs were utilized. The FC pairs differed in only one letter and were matched for frequency (Carrol, Davies & Richman, 1971). Each reader was tested for recognition of 18 each of TS, FS and 36 BS words. Words were counterbalanced across participants so that each word appeared equally as often in the TS, FS and BS conditions across readers. The priming procedure used the Spelling instructions.

*Results*

One participant was dropped for failing to respond to at least 90% of the trials on all conditions; participants were added until $B_{Acc} = 0.75$, N = 105. The Similar FC target prime, $P_{TS} = .05$; the foil prime, $P_{FS} = -.04$.

*Target Decay By FC Cues.* As in Exps. 2 & 3, $B_{Acc}$ was lower for Right than Left FC words (.74 versus .77), and given the negative accuracy skew, Left FC $P_{TS}$ was larger than Right (.06 versus .04) and $P_{FS}$ was modestly lower on the Left (-.05 versus -.04).

*Discussion*

The $P_{FS}$ of .05 is equal to the $P_{FD}$ of Exps. 2 & 3. That priming a foil similar to the target produces a larger decrement in target recognition ($P_{FS}$ = -.04) than the $P_{FD}$ = .00 of Exps. 2&3, clearly supports the C-BFAR bias model.

The significant $P_{FS}$ replicates the original Ratcliff et al. (1997) assumption that repetition priming biases target recognition when the FC alternatives are Similar. The absolute magnitude of the Similar FC primes is about half that of Ratcliff et al. (1997). Exp. 4 suggests that their larger magnitudes are not due to the prime study instructions or multiple study-test blocks but rather to the longer study-test delay intervals we used in order to minimize explicit recall of prime words at test, a delay that reduces prime magnitude (Ratcliff & McKoon, Exp. 7). Although the absolute size of the negative Similar foil prime, $P_{FS}$, is 20% smaller than the target prime, $P_{FS}$, we note that both primes vary in magnitude across several conditions, and caution against any simple conclusion about their relative size.

*Target Decay By Left-Right FC Cue Order.*

Following target presentation, however, the same word recognition pathway must be immediately re-employed to recognize the FC cue words, so this sequence requires that some record of the partly processed degraded target be held in a temporary neural buffer *outside* the bottom-up word recognition network. We assume that this temporary buffer must consist of previously learned letter features so that the lowest level of buffer may be at the Hidden F<=>L or Letters layer located in the visual word form area (VWFA) in the left ventrical occipital temporal area of the cortex (Ludersdorfer, Schurz, Richlan, Kronbichler, & Wimmer, 2013). It is highly unlikely that such a buffer has the stability of target features available on the computer screen in a conventional FC test. Following FC cue recognition, the buffered target features must be returned to the word recognition network. A diffusion model assumes that these target features are compared one at a time to the two FC alternatives (Ratcliff et al. 1997) until a decision is reached. Attractor and connectionist models assume that sets of target features are compared in parallel with the two FC alternatives.

But given the possibility that reading the left FC cue and comparing it to target features might degrade or bias the stored target features we asked whether readers might order, or even alternate, their left-to-right FC decision strategy and pursue a target <=> left FC word recognition process up to some temporary accuracy criterion before reading the left FC word and pursuing the target <=> right FC word test. Our concern was supported by a study (N=200), not reported here, which showed that $B_{Acc}$ was .05 lower for the right than left FC word. Less target information was available for right, i.e., 2nd read, relative to left, 1st read, FC decisions. Fortunately, readers do have some awareness of their target word recognition accuracy and are able to use it to determine how to exploit FC cues to assist that recognition. The validity of their accuracy estimate is demonstrated by the positive association of their confidence with recognition accuracy and negative association with RT (Durant, 2010). The most dramatic evidence that readers balance the positive and negative values of FC cues is their differential use of the left and right cues. First, they learn that reading the FC word and comparing it to available target features tends to degrade the very target features it attempts to recognize. That leaves fewer target features available for comparison with the right FC cue, lowering its potential contribution to target recognition. Given that loss, we were more than surprised to find that, except for a small number of readers, the larger their Left-Right $B_{Acc}$ decrement, the higher their mean $B_{Acc}$! Apparently estimating the negative accuracy effect of reading both FC words before matching them to target information, most participants further estimated that accuracy could be maximized by sticking with the first FC-target decision until it failed to suggest a good match and then moving on to read the 2nd, and matching it to the depleted target features. The few participants who showed no left-right $B_{Acc}$ decrement were at the bottom or top of the $B_{Acc}$ distribution. Those with the lowest target accuracy apparently didn't learn the advantage of the Left-Right strategy in the first place, so their target features were degraded by reading both FC words before the target recognition process began. Those with very highest $B_{Acc}$ had minimal left-right loss because they processed target features sufficiently well that they could be stored in a more reliable buffer – presumable in the PFC.

*General Discussion*

Prior reading of a word substantially decreases the time to accurately recognize the same word minutes or even hours later. This savings is accomplished by reducing the BU word feature information needed to accomplish accurate lexical word recognition. Fewer BU features are needed because the conceptual, cognitive features, once activated by recognition of the lexical word, remain active for some time. And during that active interval they return modest activation to the representation of the lexical word, effectively lowering the recognition threshold of the primed lexical word. That is, fewer BU features are needed to accomplish lexical recognition. We call this process of continuous anticipation of the next word, *proception*. We assume that the savings in lexical word recognition is matched by savings in the conceptual-cognitive re-*cognition* of the primed words. Understanding the processes of proception may also contribute to understanding the larger domain of prospection - processes by which the perceptual-cognitive neural networks anticipate future events (Szpunar & Schacter, in press; Szpunar, Shrikanth & Schacter, in press).

*Priming Facilitation is Accomplished By Recurrent Lexical Word Bottom-up<=>Prefrontal Cortex<=>Frontal Cortex Top-down Connections.* The BU sequence of the "public" stages of word recognition, from letter features in the occipital cortex, to letters, to words, in the left temporal lobe have long dominated traditional recognition models. But as Heeger (2017) notes, sensory processing models have consistently ignored the essential contribution of recurrent TD neural process in learning, recognition and memory. And theories that do study TD processes, tend to represent them independently of BU processes. In a recent evaluation of dual process theories of prospective memory - remembering to execute future intentions – Shelton and Scullin (2017) lament that TD and BU processes are always studied in isolation. Emphasizing the dynamic interaction of these two processes directions, they argue that identifying the interaction of TD and BU processes is critical to understanding cognition and behavior. Bidirectional connectivity is ubiquitous in the cortex (Felleman & Van Essen, 1991; Levitt, Lewis, Yoshioka

and Lund, 1993; White, 1989).  For many years, the functions of the high-density TD neurons in the visual perception system were a mystery.  But when early computation neural networks demonstrated that they were essential for both learning and recognition, their critical recognition function was acknowledged.

Following the active memory model of O'Reilly et al.  (2000) and the MRI priming studies of Buckner (2010) and Buckner et al. (2000), we assume that recurrent interactive networks link the LTC lexical word recognition representations via hub pathways with PFC and fC NNs that represent the more abstract cognitive associates of the lexical word.  Lexical recognition of a word activates the interacting PFC<=>fC NNs that represent its context-appropriate conceptual, cognitive associates -- its meaning.  We suggest that these interconnected PFC<=> fC NNs also assess the value of the conceptual, context information activated by the lexical words – what may be discarded as trivial, what should be saved as a permanent memory, and what should be kept active temporarily - for minutes or for hours as primes and working memory.  In the temporary case, these PFC<=>fC conceptual networks pass activation back down to the primed lexical words so long as the context features of primes and targets are maintained.  They are *not* "recruited" or "retrieved" by the target. They *sustain* lower lexical word thresholds so that whenever a primed target next enters the BU lexical recognition network, fewer BU features are needed for accurate recognition.  As C-BFAR demonstrates, this TD activation of a lexical word is the lexical prime "memory" - the memory that facilitates lexical recognition by lowering its lexical threshold to BU sub-lexical input.

*Repetition Priming as Learning.*  Priming unfamiliar, i.e., low frequency, words initiates a long-term learning process, which must not be confused with the processes that produce the temporary repetition prime facilitation.  Bowers (1999, 2000) and Bowers, Damian, and Havelka (2002) have demonstrated that a major effect of a single reading of the prime word is a long-lasting modification of the lexical recognition network that produces a stable increment in recognition accuracy.  Such an increment is consistent with priming of low frequency words where recognition accuracy has not yet reached an asymptote.  Ratcliff et al. (1997) used the Seidenberg and McClelland (1989) connectionist model to test this priming as learning process but did not find the predicted increment because their model had already reach its upper accuracy limit before the addition one-trial prime was applied and/or because the learning rate employed by the model was too small to produce a single trial increment.  But in an extensive reevaluation of this model, Bowers (1999, 2000) not only found empirical support for the priming as learning effect but successfully simulated it with the Seidenberg and McClelland connectionist model (Bowers et al., 2002).  Once this learning has reached its asymptote an additional single trial cannot produce an additional sensitivity increment.  The primary effect of repetition priming facilitation, therefore, is among words in the mid frequency range.  To avoid a learning confounding, the present experiments employed mean word frequency of 140/million (Carrol, et al., 1971), much higher than the lower 50/million (Kucera & Francis, 1967) reported by Bowers (Exp. 6, 1999) which showed a learning effect.

*Repetition Priming as Bias*. The failure of memory systems models to account for the facilitating effect of repetition priming on word recognition, as noted earlier, led Johnson et al. (1984) and Ratcliff et al. (1989; 1997) to ask whether this facilitation is accomplished by temporarily biasing the recognition process. Framed in terms of SDT where signal (target word) and noise (foil word) are equi-probable, bias is a shift in the recognition criterion in favor of the primed word. For e.g., following the prime, *foot*, but seeing only the features, *foo..*, of the target word, *foot*, and given the FC cues, *foot/ food*, the reader is more likely to decide on the primed alternative, *foot*. But this decision could also be attributed to an improvement in sensitivity, i.e., *d'*. Evidence for bias also requires that priming a similar, but different, word, e.g., *food,* will produce a decrement in recognizing the target, *foot*. That is what Ratcliff et al. (1997) found and we have replicated here. SDT also says that the criterion shift produced by the prime will favor the target and produce an increase in target recognition accuracy when the FC alternatives are dissimilar, e.g. *foot*/*slim*. But not finding a positive $P_{TD}$, Ratcliff et al. (1997) Exp. 1, N=16, rejected "existing models" that assume "that prior exposure to a word changes some property of the representation of the word itself. (For example,) the resting activation level for the word might be changed,.." (p 339). To solve this behavioral data problem, they proposed a diffusion model in which counters are the attractors in the lexical recognition process. But counters cannot be attractors unless they represent some cognitive or neurocognitive entity. Correcting the data problem, our experiments and the C-BFAR model suggest that the lexical NNs activated by the 2 FC cue words are the counters in the Ratcliff and McKoon model.

The positive .05 $P_{TS}$ and $P_{TD}$ Exps. 2, 3 and 5 (N=273) show that priming does, indeed, modestly elevate the activation level of the word's lexical representation, thereby lowering its response threshold to BU word feature input. This facilitation is the *positive* side of bias. With both BU and TD activation, readers may report that the primed word "jumps out" at them (Ratcliff et al., 1988). Only a few letters may suffice to "recognize" it, thereby speeding recognition. Note, that by lowering the lexical threshold for the primed word, the facilitating effect of the prime is immediate, and is not initiated by weakness or failure of the BU recognition process as suggested by some. Evidence that facilitation is produced by bias, rather than sensitivity, as defined by SDT, is the negative bias produced by priming the FC foils. The pattern of foil primes – a negative -.04 $P_{FS}$ and .0 $P_{FD}$ prime – can be produced only by a shift in the lexical decision criterion. For example, assuming the activation of the lexical representation of the primed word *spice*, has been temporary elevated, relative to the similar words, *spite, spine, spire* and *spike*. Presented with only a fraction of the target words letters, e.g., *spi*, the partly activated lexical *spice* representation will easily dominate its similar neighbors. Given all 5 letters, the *spice* RT will be faster. On the other hand, if the target is different, e.g., *spine*, the correct response will be slower, and there will be more recognition errors. And that illustrates how the positive primes, $P_{TS}$ and $P_{TD}$, produced by a SDT criterion shift in favor of the target word, qualify as "bias."

*Repetition Priming Facilitation as Context-Dependent Sensitivity*. However, the source of the "bias" differs in two critical respects from the criterion shift in traditional SDT. First, the lexical recognition networks of the similar target (signal) and foil (noise) networks are not independent of each other, but as noted above, are closely inter-related by their common letter-string input (in the Hidden L-W layer) and strong reciprocal inhibitory connections (in the Word layer).  Second, the facilitating effect of the repetition prime on target word recognition is incompatible with the standard SDT bias model where p(signal, i.e., target) = p(noise, i.e., foil). In SDT, any gain from bias in favor of the target, e.g. *died*, is erased by the cost (error) of recognizing the foil, e.g., *lied*, when it becomes the target.  By contrast, in the present paper, the similar words, *context* and *content* have similar probabilities in the *Corpus of Contemporary American English*. But after reading the word *context*, as a prime in this paper, p(*context*) in the remainder of the paper increases to 46 times that in the Corpus, while p(of the similar word, *content*) = 0.   Following the word, *bias*, in this paper, its frequency increases to 670 times that in the Corpus of American English; and following the first occurrence of the word *prime*, its frequency increases by 2,500 times, while the probability of similar words, e.g., *price, pride, prize* = 0.  Clearly, traditional SDT, where signal (correct target word) and noise (similar, but wrong word) are equi-probable is inappropriate.  Rather, this temporary prime window may be described by a *Near Noiseless Limit Signal Detection Theory* where the relative probability of the prime repetitions far exceeds that of the similar but different alternatives.

   *The Larger Context.*  How do the PFC<=>fC conceptual-context networks anticipate, i.e., predict, this increase in prime-target word probability? Livne & Bar (2016), MacLeod et al. (2000), Masson et al. (2003), Sadeghi, McClelland, & Hoffman (2014) and Schooler, Shiffrin & Raaijmakers (2001) have repeatedly demonstrated that the strength of the repetition prime is supported by an array of context features that share the prime-target space-time window.  These context features may include both the well-learned rich conceptual, cognitive and value associations of each lexical word, and the reader's external and internal environment. As Masson et al. (1990) have shown, these may, in turn, vary with the physical environment context of the reader.  For example, a newspaper account of an animal fair may sustain the strength of primes such as *prime, price, pride*, or *prize* much better than an article on politics, while moving from the animal fair to politics article may immediately decrease the strength of those primes. The final recognition processes for both words and their conceptual contexts takes place in the rich interconnected neural networks of the forebrain where recognition is coupled with frontal lobe determination of word value and the decision to discard, to hold temporarily, or to save in long-term memory (See, Antrobus, 2017; O'Reilly et al., 2000; Wamsley, 2017).

   *Restoring Cognition to Recognition*.   The TD PFC<=>fC input to the lexical word representation in the LTC demonstrates the contribution of conceptual-cognitive word "meaning'" to lexical word recognition.  As extensive evidence by Schacter, Buckner, Wig and their colleagues demonstrates, the PFC and fC TD interactions with the LTC lexical word representations suggests that all word recognition is the product of this collaboration. Lexical recognition is an intermediate, but not the final step of word recognition. The final step in word "re*cognition*" is the re-activation the conceptual-cognitive features of a word -- the activation of some abstract semantic, syntactic and other conceptual feature representations previously associated with the target word in contexts similar to the present context (See, Bergerbest et al.,

2004; Henson, 2003; Maccotta et al., 2004; Schacter, 1992, 1994; Tulving et al., 1990).  We assume that these frontal and prefrontal representations interact with each other and send activation back through the hub pathway to the ATL hub (Hoffman et al., 2018)  thereby lowering the word's lexical recognition threshold (Buckner et al., 2000) as well as activating its acoustic and articulatory word cousins, which will further support the associated frontal and prefrontal recurrent networks.  We suggest that continued recurrent interactions among these representations account for the survival of repetition priming over extended time intervals – the temporary "memory" of the prime.  Note, that the recurrent bottom-up⇔top-down interactions by which the lexical word activate its abstract cognitive meanings and they, in turn, prime the lexical word which send modest activation are similar to the recurrent bottom-up⇔top-down interactions between letters and lexical words in the lexical word recognition network. That is, the reciprocal bottom-up, top-down interactions are essential to *all* stages of recognition from the occipital to frontal cortex.  If the effect of prefrontal<=>frontal context on lexical word decision is considered as bias, so also are all reciprocal interactions in the entire C-BFAR recognition network.

*Context-Dependent Sensitivity as Short-term and Working Memory*.  Our conception of the "memory" in working memory is unfortunately constrained by our conception of memory as a discrete categorical object.  Supporting this misconception is the analogy to computer memories where the binary code for information created in one site, then saved in, and later retrieved from, another site, is identical in the two sites.  But this conception is incompatible with the distributed inter-related representation of cortical memory across neural networks located across different cortical regions.  Prefrontal cortex activity has been identified in scores of working memory studies (See, Minamoto, Tsubomi, & Osaka, 2017; Ranganath, et al. 2005; Postle, 2016, 2017).  This activation comes from interactions with the vast forebrain association networks previously activated by the lexical prime word (Cole et al., 2012; Buckner, 2010; Siegel et al., 2015), which now return activation back via a hub pathway to its lexical representation origin.  The PFC<=>fC representations of kitchen are *not* copies of the LTC NNs nets that represent the lexical kitchen. This moderate lexical activation *is the* local representation of the prime memory. This lexical memory exists at the pleasure of the "memory" of the entire supporting network, none of which corresponds to the traditional concept of a local discrete prime memory.

*The Facilitating Effect of All Context Primes, and Their Participation in Reading*.

The effect size of repetition priming on word recognition while reading newspapers and books may be much stronger than in the laboratory procedure where the prime-target context is limited to the laboratory facilities, and prime-target latency is over 15 minutes.  Furthermore, repetition prime facilitation is but one of several context primes that facilitate word recognition in everyday reading. Conceptual associations and syntax primes may have a much larger facilitation effect than repetition of the same word and may substantially facilitate recognition.  And these context primes may collaborate to substantially facilitate recognition of almost every next word.  Again, this TD PFC<=>fC context facilitation is not an additive effect, but rather an interaction with the BU input to the "public" lexical word networks, which in turn, complete conceptual word recognition in the fC word meaning networks. Rather than a BU recognition process terminating with lexical recognition, word recognition is accomplished by multi-layered interactive networks that extend from the line detectors in the occipital cortex to conceptual word meanings in the fC.

*Future Research*. Context-dependent recognition facilitation rests on the assumption that the conceptual-cognitive processes triggered by the prime word estimate increased future repetitions of the same word. To enhance the power of behavioral and fMRI studies of repetition priming it may be helpful to examine normal text reading environments to identify prime-target text contexts that show the largest increases in the frequency of primed targets.  The shared prime-target context in laboratory research is that the computer apparatus and laboratory environment, and this association may obscure the effects of text associations on prime and target. This confound may be reduced by using video goggles to manipulate the visual context of prime and target.

A critical test for context-dependent recognition facilitation is whether the duration of prime magnitude coincides with the duration of the context.  At this time there are no systematic behavioral or fMRI measures of repetition priming decay over time. If initial indices can be made sufficiently strong that they should be able to determine this relation over time.

Word recognition accuracy has provided the simplest domain for identifying the collaboration of bottom-up⇔top-down neural recognition processes. As these processes become better identified we will be able to extend them to the larger recognition domain of object recognition and the speed-accuracy recognition criteria (See, Benoni, Harari & Ullman, 2010).

## Conclusions

Recognition of every word in a newspaper, book, or email is facilitated by features of the context in which the word appears. The facilitating context may be, the conceptual meaning of prior text, its syntax, and even pictures accompanying the text or the reader's physical environment, such as one's cell phone, or computer.  But the facilitation is so complex, yet so automatic, that we tend to be unaware of its contribution and assume that word recognition is simply a bottom-up process that terminates with lexical recognition.  Given that complexity, context research has proceeded on the assumption that the analysis should start with the the simplest possible form of word recognition context, namely, repetition priming, the effect of prior reading of exactly the same word.

That prior recognition of the same word, the prime, facilitates recognition of a familiar target word is has been demonstrated by several hundred studies. They conclude that facilitation is accomplished by lowering target recognition threshold.  Although several models suggested that the lowered threshold is produced by the creation, and subsequent retrieval, of a short-term, hippocampus-created, prime memory, neurological evidence eventually rejected those models. This led Ratcliff and McKoon to propose that the facilitation is not recognition sensitivity, but bias, as represented in SDT.  The five repetition priming experiments in this paper, N=274, support their bias assumption.

SDT represents bias as a shift in the decision criterion. But every "word" is represented in neural networks that span the cortex, from letter features in the occipital cortex, to the lexical word in the left temporal lobe, to its conceptual representations in the frontal cortex. Priming theories differed as to whether the priming effect occurs at the letter level or at the lexical word level. Therefore, we have constructed a neural network model, Context-Biased Fast Accurate Recognition (C-BFAR) that represents word recognition across six layers from letter features to lexical words. We find that the bias pattern of behavior priming data is matched only when the priming produces modest activation in the neural nets that represent the lexical word. By interacting laterally with similar lexical words, and downwards with each of their sub-lexical letter sets, this activation lowers target word threshold when the target is primed, and increases its threshold when a similar, but different, word is primed. We assume that this modest activation to the primed lexical representation comes from the conceptual-cognitive networks in the frontal and prefrontal cortices. By lowering the lexical word threshold to the primed word and increasing threshold of a similar, but different, word, this activation, in effect, produces the SDT recognition criterion shift that biases recognition.

A critical repetition priming fMRI study of Buckner et al. (2000) confirmed this assumption. They found that the activation of the lexical word's representation in the left temporal and parietal lobe regions not only moves forward/up to activate regions of the prefrontal and left frontal cortex networks that represent the conceptual, cognitive meaning of the lexical word, but those regions then pass activation back down to their source in the temporal and parietal lobes. This pass-back from the prefrontal to left temporal and parietal regions supports the C-BFAR network model where the conceptual cognitive meanings of a word contribute to the lexical recognition of subsequent words. By representing the meaning of one word they constitute the context for the lexical recognition of subsequent words. In the case of repetition priming, the cognitive, conceptual meaning of the prime becomes the context that facilitates recognizing of subsequent lexical target words. This top-down prefrontal context activation identified by Buckner et al. appears to be the source of this fast accurate recognition bias. The final step of target word re*cognition* occurs when the primed lexical target representation sends activation back up to activate the appropriate prefrontal<=>frontal conceptual cognitive networks. Cognitive context representations in the frontal<=>prefrontal cortex actively contribute to the lexical recognition of every word we read. This reminds us that "public" recognition step, i.e., *lexical* recognition, is but an intermediate step in re*cognizing* the cognitive meaning of a word. Since traditional bottom-up recognition models generally terminate at the lexical level they represent only part of the recognition process.

Given that there are at least as many descending as ascending neurons in the word recognition system, word recognition is the collaboration of bottom-up, top-down neural net computing that extends from the occipital to frontal cortices. From this perspective, each network layer creates contexts that assist neural decisions in the layer below it. For e.g., lexical words provide contexts for sub-word letter sets, which in turn, provide contexts for, i.e., bias, computing letter decisions. These top-down biases reduce the neural computing load at each layer, so the accumulated saving substantially enhances cortical efficiency.

If the prime and a similar foil were equiprobable, any gain from bias in favor of the target would be offset by an equal loss from misrecognizing the similar foil.  However, informal analyses of printed texts suggest that the prime initiates a temporary time-space context window within which the probability of a small set of words or objects is substantially higher than for the same words or objects outside of that context. As noted earlier, the first occurrence of the word *prime* in this paper is followed by its repetition more than a thousand times its frequency in the Corpus of American English, while the probability of all similar (potential foil=wrong) words = 0.  Again, we need to bring cognition back to recognition.  Reading the prime includes recognizing its conceptual-cognitive context.  If that context says that the probabilities of similar, but different, words are close to 0.0, then the context of the prime word temporarily acts like a feature of the word itself, thereby paradoxically converting the prime bias into *de facto* sensitivity.

Repetition priming is but one of many classes of priming, e.g., semantic, syntactic, visual, and auditory. Together they create the continuously changing contexts of all waking perception that collaborate with bottom-up sensory input to substantially speed recognition -- without sacrificing accuracy. That is, every second of waking perception occurs in a context, many of whose words and objects are familiar. By lowering the recognition thresholds of the lexical representation in the temporal cortex, the frontal and prefrontal associates of a word speed recognition by anticipating the likelihood of the next word. Over an entire day, the cumulative savings in time of this anticipation is large. We coin this anticipatory process, "proception."  Unfortunately, the positive speed-accuracy payoff of this bias flies in the face of the largely pejorative use of the word "bias".  Popular usage of "bias" implies negative race, ethnic or gender bias - judgments of one's superiority over people in other social classes. That these biases are inaccurate and destructive promotes the popular view that all bias is "bad" and should be corrected.

Antrobus et al.	Recognition Sensitivity in Contexts

Table 1

Summary of Procedures

Priming Conditions

\*\* Prime \*\*      \*\*\*\*\*\*\*\*\*\*\*\*\*\*\*\* Test \*\*\*\*\*\*\*\*\*\*\*\*\*\*\*\*\*\*\*\*

| Condition | Target | Mask | FC Pair | Prime |
|---|---|---|---|---|
| Similar FC Word Pairs [1] | | | | |
| TS | [ **died** ] | [died] | @ @ @ @ | [ died/lied ] | $P_{TS}$ = TS-BS |
| FS | [ **lied** ] | [died] | @ @ @ @ | [ died/lied ] | $P_{FS}$ = FS-BS |
| BS | -- | [died] | @ @ @ @ | [ died/lied ] | |
| Dissimilar FC Word Pairs [2] | | | | |
| TD | [ **died** ] | [died] | @ @ @ @ | [ died/soon ] | $P_{TD}$ = TD-BD |
| FD | [ **soon** ] | [died] | @ @ @ @ | [ died/soon ] | $P_{FD}$ = FD-BD |
| BD | -- | [died] | @ @ @ @ | [ died/soon ] | |

[1] In Experiments 1, & 4        [2] In Experiment 1, 2 & 3

Table 2

Baseline and Priming Values: Dissimilar FC Experiments 2 and 3, Similar FC Experiment 5, And C-BFAR Model

Experiments

| Dissimilar FC | | Similar FC | |
|---|---|---|---|
| 2 (N = 72) | | 5 (N = 105) | |
| $B_D$ | .75 | $B_S$ | .75 |
| $P_{TD}$ | .05 | $P_{TS}$ | .05 |
| $P_{FD}$ | -.02 | $P_{FS}$ | -.04 |
| 3 (N = 97) | | | |
| $B_D$ | .75 | | |
| $P_{TD}$ | .05 | | |
| $P_{FD}$ | .02 | | |
| 2 + 3 (N = 169) | | | |
| $B_D$ | .75 | | |
| $P_{TD}$ | .05 | | |
| $P_{FD}$ | .00 | | |

Recognition Sensitivity in Contexts

| $B_D$ | .75 | $B_S$ | .75 |
|---|---|---|---|
| $P_{TD}$ | .05 | $P_{TS}$ | .05 |
| $P_{FD}$ | -.006 | $P_{FS}$ | -.04 |

Figures

Figure 1.  Recognition Sensitivity in Contexts (RSIC): Neural Network Schematic

John Antrobus, Department of Psychology, City College, City University of New York, NY; Yusuke Shono, Department of Psychiatry and Behavioral Sciences, University of Washington School of Medicine, Seattle, WA; Wolfgang M. Pauli, Artificial Intelligence Platform, Microsoft, Redmond, WA; Bala Sundaram, Department of Physics, and Graduate Studies, University of Massachusetts, Boston, MA.

This research was supported by several PSC-CUNY awards to the senior author.

[1] C-BFAR produces a Dissimilar Foil prime of -.006, just a tad larger than the behavioral prime of 0.00.  This suggests in the word learning phase, C-BFAR the word network acquired a very small amount of inhibition between most or all words.  This may be a consequence of the small word set of 150 words. On the other hand, the Dissimilar Foil primes values varied from sample to sample so we must simply conclude that both behavioral and C-BFAR Similar Foil primes are very close to 0.0.

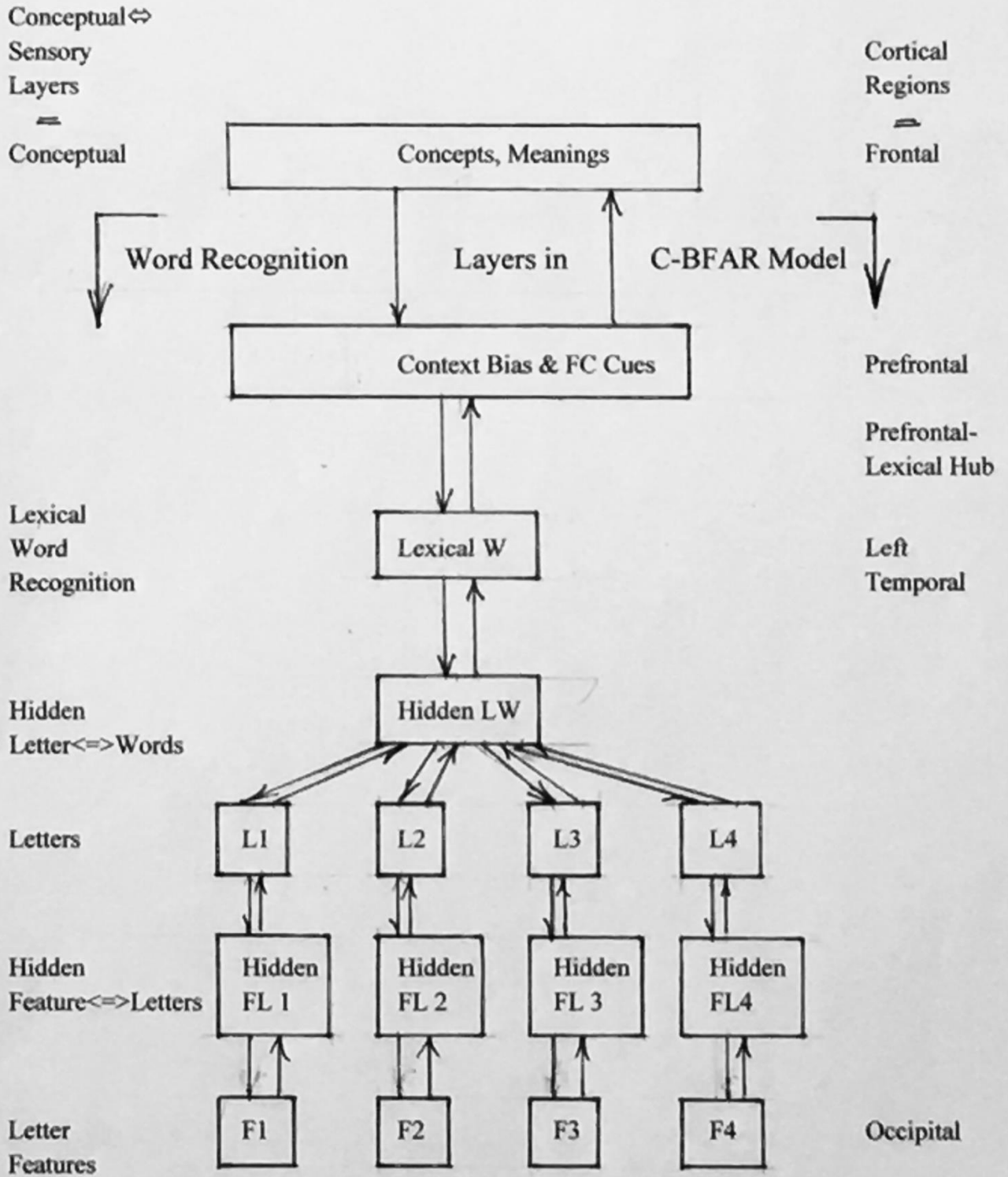